\documentclass[aps,preprint]{revtex4-2}
\usepackage{amssymb}
\usepackage{amsmath}
\usepackage{graphicx}
\usepackage{bm}
\usepackage{physics}
\usepackage{hyperref}
\usepackage{float}
\usepackage{subcaption}
\usepackage{dcolumn}
\usepackage{caption}
\usepackage{natbib}
\usepackage{xcolor}
\usepackage{overpic}

\begin{document}

\title{Frequency Heterogeneity can Promote Order yet Undermine Stability in Kuramoto Networks with Higher-Order Interactions}

\author{Zheng Wang}
 \affiliation{State Key Laboratory of Mechanics and Control for Aerospace Structures, College of Aerospace Engineering, Nanjing University of Aeronautics and Astronautics, Nanjing 210016, China}

\author{Jinjie Zhu}
\email{jinjiezhu@nuaa.edu.cn}
\affiliation{State Key Laboratory of Mechanics and Control for Aerospace Structures, College of Aerospace Engineering, Nanjing University of Aeronautics and Astronautics, Nanjing 210016, China}

\author{Wenchang Qi}
 \affiliation{State Key Laboratory of Mechanics and Control for Aerospace Structures, College of Aerospace Engineering, Nanjing University of Aeronautics and Astronautics, Nanjing 210016, China}

\author{Xianbin Liu}
\email{xbliu@nuaa.edu.cn}
\affiliation{State Key Laboratory of Mechanics and Control for Aerospace Structures, College of Aerospace Engineering, Nanjing University of Aeronautics and Astronautics, Nanjing 210016, China}

\begin{abstract}
We investigate the interplay between frequency heterogeneity and higher-order triadic interactions in a ring network of Kuramoto oscillators. While both factors individually disrupt ordered states, their combination produces unexpected collective behavior. In the strong triadic coupling regime, moderate frequency heterogeneity substantially increases the global order parameter, with an optimal heterogeneity strength growing approximately linearly with triadic coupling strength. Basin stability analysis reveals that this order-promoting effect arises from a global restructuring of the attractor landscape: frequency heterogeneity shifts the attractor competition in favor of more ordered configurations. Linear stability analysis of frequency-locked twisted states reveals a competing effect: frequency heterogeneity monotonically erodes linear stability and reduces the probability of frequency locking. These two competing mechanisms, basin enlargement and linear destabilization, together account for the non-monotonic dependence of the order parameter on heterogeneity strength. Our results demonstrate that frequency heterogeneity can play a constructive role in oscillator networks with higher-order interactions.
\end{abstract}

\maketitle

\section{Introduction}
\label{sec:intro}

Synchronization and collective order are central themes in the study of coupled oscillator networks, with applications ranging from power grids and neural circuits to biological rhythms and coupled lasers~\cite{WINFREE1967,kurths2001,istvan2002,Arenas2008,Florian2013,zhu2025complex}. The Kuramoto model and its variants have long served as the canonical framework for understanding how local coupling gives rise to global coherence in large populations of phase oscillators~\cite{kuramoto1984,Strogatz2000,Acebr2005}. Despite its apparent simplicity, the model exhibits a remarkably rich phenomenology, including continuous and discontinuous transitions to synchrony, multistability, and a variety of partially ordered states~\cite{Abrams2004,wiley2006,Erik2013,skardal2019}. These features have motivated numerous extensions of the original model to incorporate more realistic aspects of real-world oscillator populations.

Two such extensions have attracted considerable attention  recently. The first is the introduction of higher-order interactions, which couple three or more oscillators simultaneously in a way that cannot be decomposed into pairwise terms~\cite{Battiston2020,Battiston2021,Bick2023,Boccaletti2023}. Such nonpairwise couplings arise naturally in neural circuits~\cite{Giusti2016,Parastesh2022}, ecological communities~\cite{Grilli2017}, and social contagion processes~\cite{iacopini2019simplicial}, and have recently been incorporated into oscillator network models to capture beyond-pairwise coupling effects~\cite{Tanaka2011,Lucas2020,skardal2020higher,Gambuzza2021}. In oscillator networks, triadic interactions have been shown to enrich the dynamical repertoire by promoting chimera states and disordered attractors~\cite{Bick2016,Millan2020,Kundu2022higher,Muolo2024phase}, and to simultaneously enhance the linear stability and stochastic stability of ordered states while reducing their basins of attraction~\cite{zhang2024deeper,wang2025how,wang2026moderate,muolo2026higher}. The second extension is the inclusion of frequency heterogeneity, the ubiquitous mismatch in natural frequencies among real oscillators. In the classical Kuramoto setting, frequency heterogeneity is well established as an obstacle to synchronization, reducing the order parameter and raising the critical coupling threshold~\cite{Sakaguchi1986,Strogatz2000,Francisco2016}.

Yet the perception that disorder is necessarily detrimental to collective order has been challenged by a number of studies. It was shown early on that spatial disorder can tame spatiotemporal chaos in arrays of coupled oscillators, giving rise to more regular collective patterns~\cite{braiman1995taming}. More recently, it has been shown that in oscillator networks with communication delays, random frequency heterogeneity can stabilize otherwise unstable synchronization states, with intermediate levels of heterogeneity outperforming purposely designed parameter assignments~\cite{zhang2021random}. These results share a common theme: disorder can restructure the dynamical landscape in ways that favor coherence. However, the mechanisms differ substantially across systems, and it remains unclear whether analogous effects arise in oscillator networks with higher-order interactions, where the attractor landscape is fundamentally more complex.

A natural question then arises: when higher-order interactions and frequency heterogeneity are present simultaneously, do their individually disruptive effects simply compound, or can something more unexpected emerge? This question is nontrivial because higher-order interactions fundamentally alter the attractor landscape by creating new competing attractors and reshaping existing basins, and it is not clear a priori how frequency heterogeneity navigates this modified landscape. Unlike the delay-induced stabilization mechanism of~\cite{zhang2021random}, which operates through the linear stability of a synchronization orbit, any order-promoting effect in our setting would have to arise from a fundamentally different route, namely the global restructuring of basins of attraction.

In this paper, we address this question by studying a ring network of Kuramoto oscillators with pairwise and triadic interactions, and Gaussian-distributed natural frequencies. We systematically map the global order parameter across the full parameter space of triadic coupling strength and frequency heterogeneity, and elucidate the underlying mechanisms through basin stability analysis and linear stability analysis of frequency-locked twisted states. Our results reveal a nontrivial interplay between frequency heterogeneity and higher-order interactions, in which two individually disruptive factors combine to produce unexpected collective behavior through global basin restructuring.

The remainder of the paper is organized as follows. Section~\ref{sec:model} introduces the model and defines the order parameters. Section~\ref{sec:order} documents the effect of frequency heterogeneity on the global order parameter across the full parameter space. Sections~\ref{sec:basin} and~\ref{sec:linear} provide mechanistic explanations from the basin stability and linear stability perspectives, respectively. Section~\ref{sec:conclusion} summarizes the findings and outlines directions for future work.

\section{Model}
\label{sec:model}
We consider a ring network of $n$ coupled phase oscillators, where each oscillator $i$ interacts with its $r$ nearest neighbors on each side. The dynamics is governed by a generalized Kuramoto model incorporating both pairwise and triadic interactions~\cite{zhang2024deeper,wang2025how,wang2026moderate}:
\begin{equation}
    \dot{\theta}_{i} = \omega_{i}
    + \frac{\sigma}{2r} \sum_{\substack{j \in \mathcal{N}_{i}}}
      \sin(\theta_{j} - \theta_{i})
    + \frac{\sigma_{\Delta}}{2r(2r-1)}
      \sum_{\substack{j \in \mathcal{N}_{i}}} \sum_{\substack{k \in \mathcal{N}_{i} \\ k \neq j}}
      \sin(\theta_{j} + \theta_{k} - 2\theta_{i}),
    \label{eq:model}
\end{equation}
where $\mathcal{N}_{i} = \{i-r,\ldots,i-1,i+1,\ldots,i+r\}$ (indices taken modulo $n$) denotes the neighbor set of oscillator $i$, and the condition $k \neq j$ in the double sum ensures that all three oscillators involved in each triadic term are distinct. Here $\theta_{i} \in [-\pi,\pi)$ is the phase of oscillator $i$, and $\omega_{i}$ is its natural frequency drawn independently from a Gaussian distribution with mean $\bar{\omega} = 0$ and standard deviation $\omega_{std}$. The parameters $\sigma > 0$ and $\sigma_{\Delta} > 0$ control the strengths of pairwise and triadic coupling, respectively, and the normalization
factors $2r$ and $2r(2r-1)$ provide proper normalization based on connection counts, keeping both interaction terms consistently scaled with the number of neighbors.

\begin{figure}[htbp]
\centering
\includegraphics[width=0.7\textwidth]{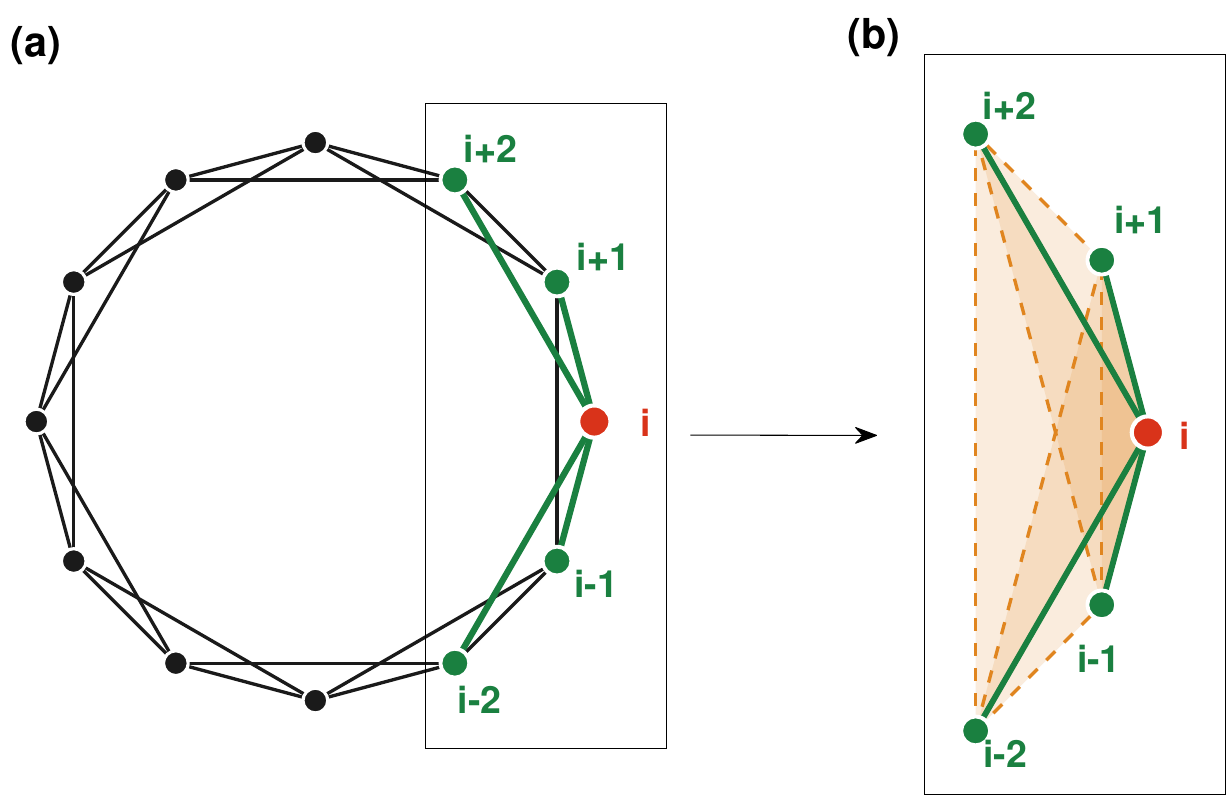}
\caption{Schematic of the ring network ($n=12$, $r=2$). (a) Global topology with pairwise interactions: green lines highlight
the couplings between oscillator $i$ (red) and its four nearest neighbors
(green nodes); black lines show remaining network edges. (b) Triadic interactions acting on oscillator $i$: each shaded triangle shares $i$ as a vertex, with $j$ and $k$ being two distinct neighbors; there are $\binom{2r}{2}=6$ such triangles in total. Green solid lines indicate pairwise coupling; orange dashed lines and shading indicate triadic coupling.}
\label{fig:model}
\end{figure}

The network topology and the two types of coupling are illustrated in Fig.~\ref{fig:model}. The pairwise term $\sin(\theta_{j} - \theta_{i})$ represents the classical
nearest-neighbor Kuramoto interaction (Fig.~\ref{fig:model}(a)), which tends to synchronize adjacent oscillators.
The triadic term $\sin(\theta_{j} + \theta_{k} - 2\theta_{i})$ is a higher-order interaction (Fig.~\ref{fig:model}(b)), in which two neighbors $j$ and $k$ jointly influence oscillator $i$. The coupling range $r = 2$ is the minimum value at which closed triangles appear in the network (e.g., the triplet $(i,\,i{-}1,\,i{+}1)$); note, however, that not every triadic term in Eq.~\eqref{eq:model} corresponds to a closed triangle. Importantly, the introduction of triadic coupling enriches the dynamical
repertoire of the system: beyond the classical synchronized state, the network can exhibit chimera states, in which spatially coexisting coherent and incoherent domains emerge, as well as fully disordered states~\cite{Bick2016,Millan2020,Kundu2022higher,Muolo2024phase}.

To quantify the collective state of the network, we define a local order parameter for each oscillator~\cite{zhang2024deeper,wang2025how,wang2026moderate},
\begin{equation}
    R_{i} = \left| \frac{1}{2r+1}
             \sum_{|l-i| \leq r} e^{\mathrm{i}\theta_{l}} \right|,
    \label{eq:local_R}
\end{equation}
where $l$ runs over all oscillators within distance $r$ of oscillator $i$, and $\mathrm{i} = \sqrt{-1}$ denotes the imaginary unit, and a global order parameter~\cite{zhang2024deeper,wang2025how,wang2026moderate}
\begin{equation}
   R = \frac{1}{n} \sum_{i=1}^{n} \mathbb{I}(R_{i} \geq \rho_c),
    \label{eq:global_R}
\end{equation}
where $\mathbb{I}(\cdot)$ is the indicator function that equals $1$ if the condition is satisfied and $0$ otherwise, and $\rho_c = 0.85$ is a threshold separating locally ordered from locally disordered oscillators. The threshold $\rho_c = 0.85$ is chosen following previous studies~\cite{zhang2024deeper,wang2025how,wang2026moderate}, balancing two competing risks: a value too low incorrectly labels disordered regions as coherent, while a value too high misidentifies higher-winding twisted states as disordered. $R = 1$ corresponds to a fully ordered twisted state, in which the phases are linearly distributed around the ring: $\theta_{i}^{(q)} = 2\pi q i/n + C$, where $C$ is an arbitrary constant
reflecting the rotational symmetry of the system and the integer $q$ is called the winding number. The special case $q = 0$ reduces to the global synchronization state in which all oscillators share the same phase. $R = 0$ corresponds to a fully disordered state, and intermediate values indicate partial order, including chimera states. Representative phase profiles of the three collective states (twisted, chimera, and disordered) are shown in the middle column of Fig.~\ref{fig:order}, illustrating how different values of $R$ reflect qualitatively distinct collective dynamics.

Following previous studies, we set $n = 83$ and $r = 2$.
This system size has been shown to be sufficient for capturing collective network dynamics, and for this system, finite-size effects have been verified to not qualitatively alter the main results~\cite{zhang2024deeper,wang2025how,wang2026moderate}. Throughout this work we fix $\sigma = 1.0$ and $\bar{\omega} = 0$, and all numerical simulations use a first-order Euler scheme with time step $\Delta t = 0.01$. For all measurements of $R$ in the presence of frequency heterogeneity, the system is first integrated for $500$ time units to eliminate transients, and $R$ is then computed as a time average over the subsequent $100$ time units. Both durations have been verified to be sufficiently long~\cite{wang2025how,wang2026moderate}.

\section{Frequency Heterogeneity Promotes Order}
\label{sec:order}

Having introduced the model, we now systematically investigate how frequency heterogeneity affects the degree of order in the system.

\begin{figure}[htbp]
\centering
\includegraphics[width=0.95\textwidth]{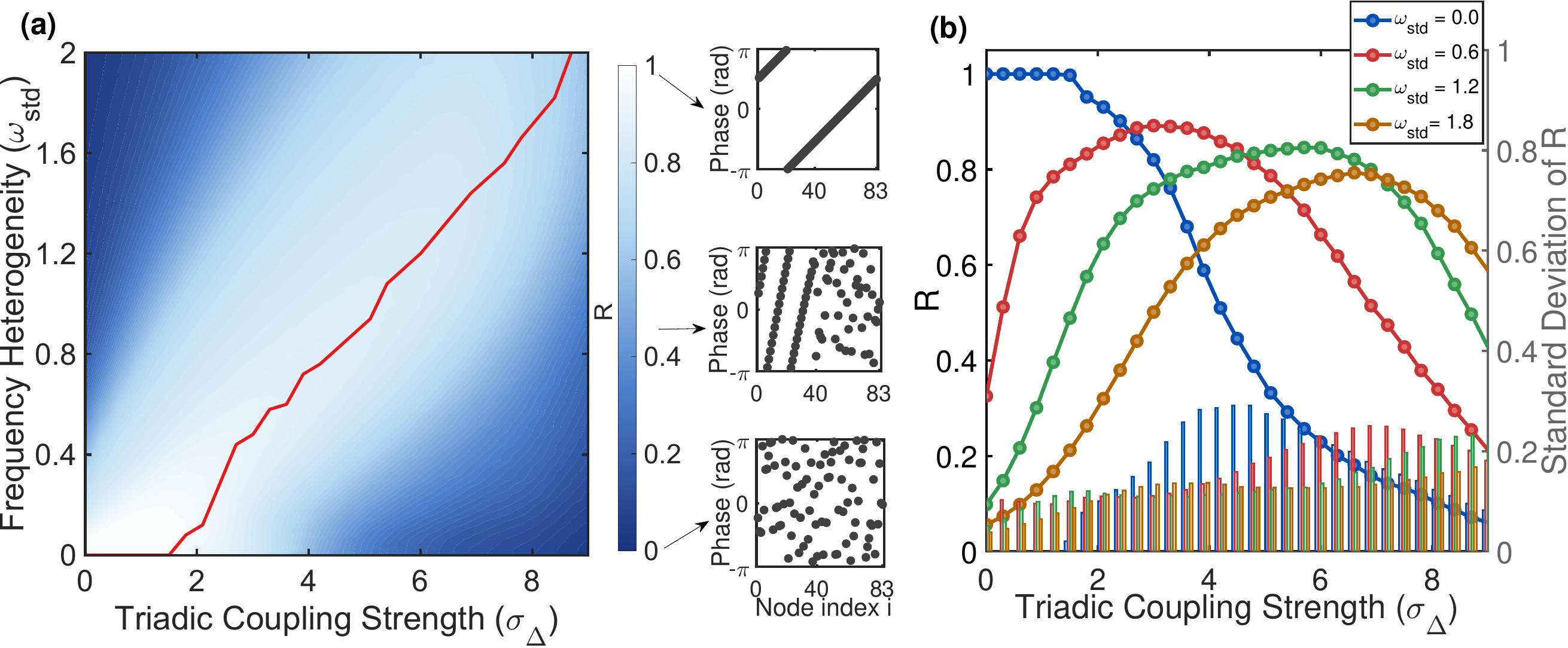}
\caption{Effect of frequency heterogeneity on the global order parameter $R$. (a) Heat map of $R$ in the $(\sigma_{\Delta}, \omega_{std})$ parameter space. The red curve marks the optimal heterogeneity $\omega_{std}^{*}$ that maximizes $R$ at each $\sigma_{\Delta}$. Insets show representative phase profiles of the three collective states: twisted state (top), chimera state (middle), and disordered state (bottom). (b) Cross-sectional curves of $R$ versus $\sigma_{\Delta}$ at four values of $\omega_{std}$. Bar chart shows the standard deviation of $R$ across independent realizations (right axis). Parameters: $n = 83$, $r = 2$, $\sigma = 1.0$, with $16{,}000$ independent realizations.}
\label{fig:order}
\end{figure}

We begin by mapping the global order parameter $R$ across the full parameter space $(\sigma_{\Delta}, \omega_{std})$. The heat map in Fig.~\ref{fig:order}, averaged over $16{,}000$ independent realizations with random initial conditions, reveals a rich dependence on both parameters. In the homogeneous case ($\omega_{std} = 0$), the system resides in a fully ordered twisted state ($R = 1$) for small triadic coupling $\sigma_{\Delta}$; as $\sigma_{\Delta}$ increases, the triadic coupling progressively disrupts the ordered structure, and chimera states and disordered states emerge and come to dominate, with $R$ decreasing monotonically. Representative phase profiles of these three collective states are shown in the middle column of Fig.~\ref{fig:order}.

Upon introducing frequency heterogeneity ($\omega_{std} > 0$), two noteworthy features emerge. First, in the small-$\sigma_{\Delta}$ regime, increasing $\omega_{std}$ further reduces $R$, meaning that frequency heterogeneity accelerates the destruction of the ordered structure, which is consistent with the classical picture in which heterogeneity hinders synchronization~\cite{kuramoto1984,Strogatz2000,Acebr2005}. Notably, within an appropriate range of $\sigma_{\Delta}$, the twisted state occupies the entire phase space as the sole attractor, and the triadic coupling does not alter this basin structure but enhances its stochastic stability~\cite{wang2026moderate}, enabling the system to withstand stronger frequency heterogeneity. As a result, the critical $\omega_{std}$ required to significantly disrupt the ordered structure increases with $\sigma_{\Delta}$, manifested in the heat map as an expansion of the ordered region toward the upper right, a trend also reflected in the slope of the red curve.

Second, in the large-$\sigma_{\Delta}$ regime where chimera and disordered states dominate the homogeneous dynamics, moderate frequency heterogeneity is able to substantially increase $R$, giving rise to a counterintuitive phenomenon in which frequency heterogeneity promotes order. The red curve marks the optimal heterogeneity $\omega_{std}^{*}$ that maximizes $R$ at each $\sigma_{\Delta}$; it grows approximately linearly with $\sigma_{\Delta}$, revealing an intrinsic relationship between triadic coupling strength and optimal heterogeneity. Once $\omega_{std}$ exceeds $\omega_{std}^{*}$, $R$ decays rapidly, indicating that excessively strong heterogeneity ultimately destroys the twisted state.

A more quantitative picture emerges from the cross-sectional curves of $R$ versus $\sigma_{\Delta}$ at four representative values of $\omega_{std}$ as shown in Fig.~\ref{fig:order}(b); the bar chart corresponds to the standard deviation of $R$ across independent realizations (right axis). For $\omega_{std} = 0.0$, $R$ decreases monotonically with $\sigma_{\Delta}$. For $\omega_{std} = 0.6$, $1.2$, and $1.8$, the curves exhibit non-monotonic behavior, each attaining a peak at intermediate $\sigma_{\Delta}$ that exceeds the corresponding homogeneous value, directly and quantitatively confirming the order-promoting role of frequency heterogeneity. As $\omega_{std}$ increases, the peak shifts toward larger $\sigma_{\Delta}$, reflecting a competition between triadic coupling and frequency heterogeneity: stronger heterogeneity requires stronger triadic coupling to shift the system toward a more ordered state. Meanwhile, the peak height decreases progressively with $\omega_{std}$, indicating that the order-promoting effect of frequency heterogeneity has an upper bound --- although triadic coupling can partially compensate for the desynchronizing effect of heterogeneity, this compensatory capacity is gradually exhausted as heterogeneity grows, and the maximum attainable degree of order declines accordingly. The standard deviation is markedly larger near the transition region, indicating heightened sensitivity to initial conditions and the coexistence of multiple stable states in this parameter regime.

Taken together, these results demonstrate that frequency heterogeneity plays a dual role: in the weak-coupling regime dominated by the twisted state, heterogeneity accelerates the breakdown of order; in the strong-coupling regime dominated by chimera and disordered states, moderate heterogeneity restores order, with an optimal $\omega_{std}^{*}$ that maximizes $R$. To understand how frequency heterogeneity influences the system, we examine its effects from two complementary perspectives: basin stability and linear stability, which are addressed in the following sections.

\section{Basin Stability Analysis}
\label{sec:basin}

The results of the preceding section demonstrate that frequency heterogeneity can substantially enhance the degree of order in the strong-coupling regime. However, the microscopic mechanism underlying this order-promoting effect remains unclear. In this section, we investigate how frequency heterogeneity reshapes the attractor competition, focusing on the evolution of initially disordered configurations and the overall effects on the system across arbitrary initial conditions.

\begin{figure}[htbp]
\centering
\includegraphics[width=1\textwidth]{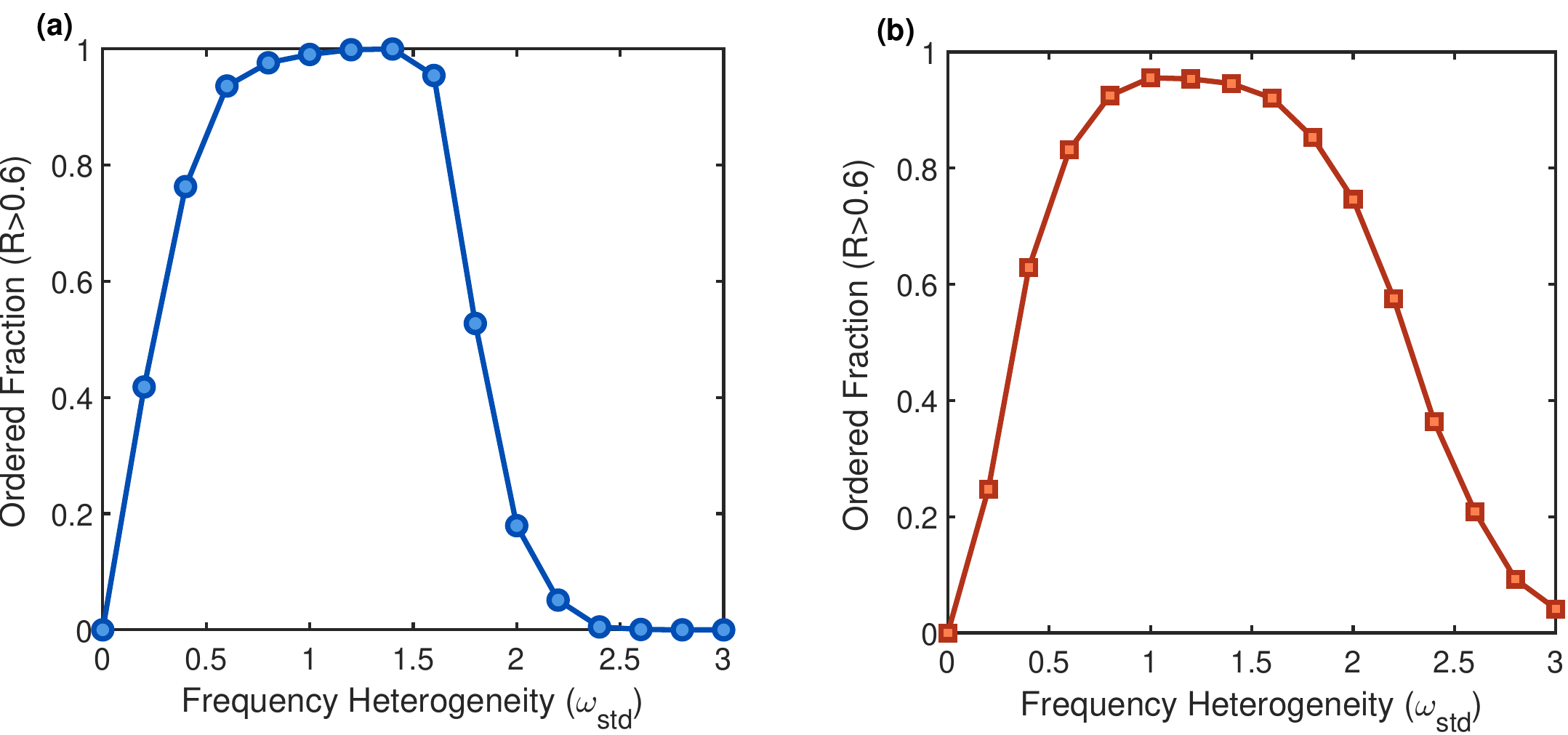}
\caption{Ordered fraction ($R > 0.6$) as a function of frequency heterogeneity $\omega_{std}$, measured from low-order initial conditions ($R < 0.6$). (a) Fixed frequency distribution shape, $1{,}600$ different initial conditions. (b) Fixed initial condition, $1{,}600$ independent frequency realizations. Parameters: $n = 83$, $r = 2$, $\sigma = 1.0$, $\sigma_{\Delta} = 5$.}
\label{fig:chimera}
\end{figure}

A natural question is whether this order-promoting effect persists when trajectories start from low-order initial conditions. We therefore compute the ordered fraction --- defined as the fraction of trajectories starting from low-order initial conditions ($R < 0.6$) that eventually reach $R > 0.6$ --- at $\sigma_{\Delta} = 5$ as a function of $\omega_{std}$ (Fig.~\ref{fig:chimera}). In Fig.~\ref{fig:chimera}(a), a single frequency distribution shape is fixed while $1{,}600$ different low-order initial conditions are sampled; in Fig.~\ref{fig:chimera}(b), a single low-order initial condition is fixed while $1{,}600$ independent frequency realizations are drawn. Both panels exhibit a consistent non-monotonic dependence on $\omega_{std}$: at $\omega_{std} = 0$, the ordered fraction is zero, indicating that in the homogeneous system low-order initial conditions cannot spontaneously evolve toward a high-order state; as $\omega_{std}$ increases, the ordered fraction rises rapidly and reaches a peak near $\omega_{std} \approx 1.0$--$1.5$, before declining monotonically to zero. This demonstrates that moderate frequency heterogeneity redirects trajectories that would otherwise converge to low-order attractors toward the high-order state, effectively shifting the attractor competition in favor of more ordered configurations. The close agreement between Fig.~\ref{fig:chimera}(a) and Fig.~\ref{fig:chimera}(b) further indicates that this effect is robust with respect to both the choice of initial condition and the specific realization of the frequency distribution.

\begin{figure}[htbp]
\centering
\includegraphics[width=0.7\textwidth]{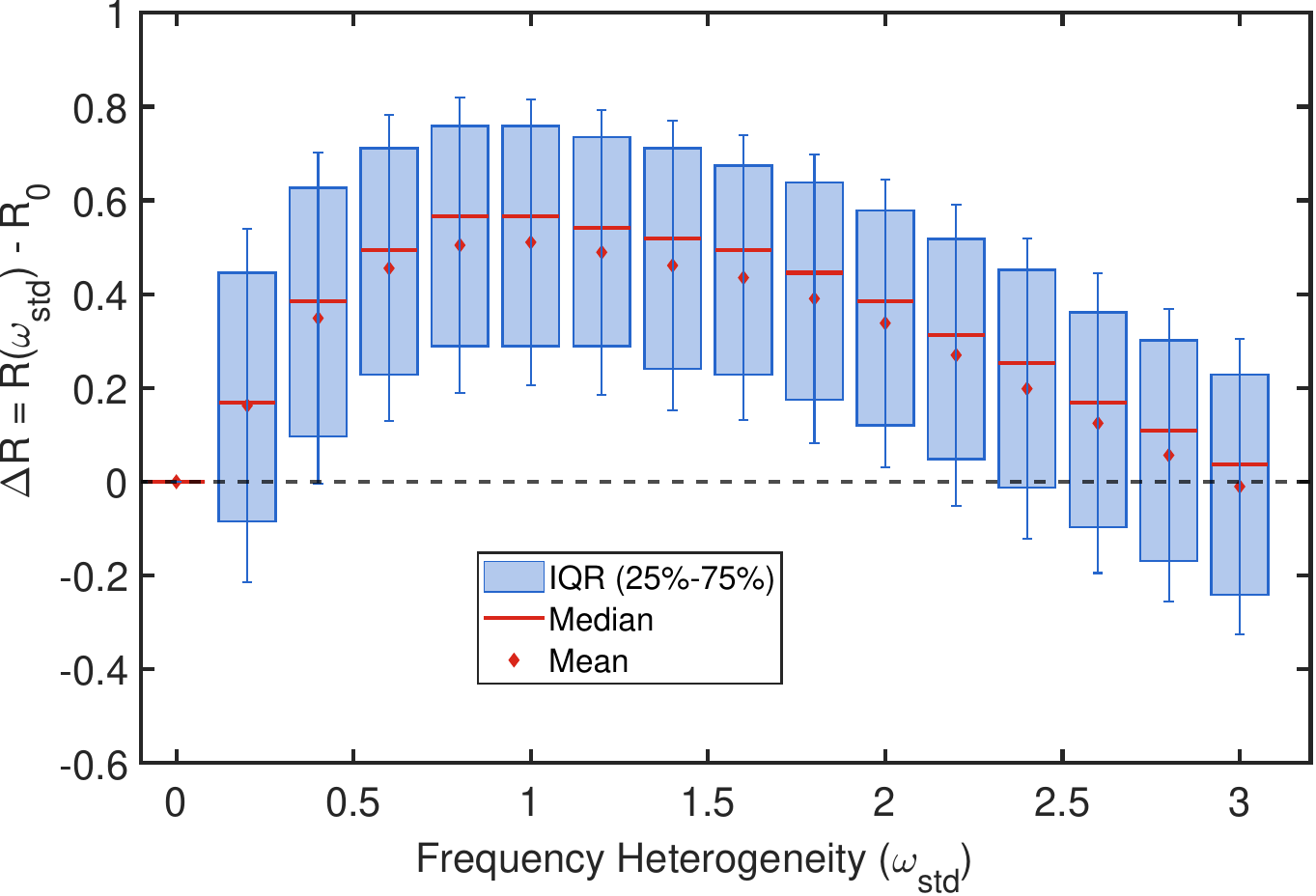}
\caption{Change in order parameter $\Delta R = R(\omega_{std}) - R_0$ as a function of $\omega_{std}$, where $R_0 = R(0)$ is the order parameter of the same initial condition in the homogeneous system. Box plots show the interquartile range (IQR, 25\%--75\%), median (red line), and mean (red point) across $1{,}600$ independent random samples at each $\omega_{std}$. Error bars indicate the standard deviation. The dashed line marks $\Delta R = 0$. Parameters: $n = 83$, $r = 2$, $\sigma = 1.0$, $\sigma_{\Delta} = 5$.}
\label{fig:xiang}
\end{figure}

The generality of this effect across arbitrary initial conditions is further assessed through the box plots of Fig.~\ref{fig:xiang}, which show $\Delta R = R(\omega_{std}) - R_0$, where $R(\omega_{std})$ is the time-averaged order parameter under frequency heterogeneity $\omega_{std}$ and $R_0 = R(0)$ is the corresponding value in the homogeneous case, measured from the same initial condition. Each value of $\omega_{std}$ is represented by $1{,}600$ independent random samples. Across all values of $\omega_{std}$, the mean and median of $\Delta R$ first rise and then fall, attaining their maximum near $\omega_{std} \approx 1.0$, consistent with the optimal heterogeneity identified in Fig.~\ref{fig:order}(a). The lower quartile remains positive up to $\omega_{std} \approx 2.0$, indicating that frequency heterogeneity promotes order for the majority of trajectories in this range. At large $\omega_{std}$, the distribution shifts downward: the lower quartile drops below zero and the mean approaches zero, indicating that excessively strong heterogeneity suppresses order for a growing fraction of trajectories. Taken together, these results statistically confirm the generality of the order-promoting effect of frequency heterogeneity, while also revealing its limitations when heterogeneity becomes too strong.

\section{Linear Stability Analysis}
\label{sec:linear}

The basin stability analysis of the preceding section reveals that moderate frequency heterogeneity shifts the attractor competition in favor of more ordered configurations, redirecting trajectories that would otherwise converge to disordered or chimera states toward more ordered configurations. However, a complementary question concerns the linear stability of the ordered states themselves. We therefore focus on twisted states, which admit exact solutions and allow a systematic linear stability analysis. In the presence of frequency heterogeneity, the exact twisted state is no longer a solution; instead, the system may settle into a frequency-locked state, in which all oscillators rotate at a common effective frequency while their phase configuration remains close to that of a twisted state. Specifically, we compute the maximum transverse Lyapunov exponent (MTLE) of such frequency-locked states in the vicinity of the $q$-twisted state for $q = 0, 1, 2$, and examine how both the existence and stability of these locked states change with $\omega_{std}$.

In the homogeneous system ($\omega_i = \bar{\omega}$), the $q$-twisted state
\begin{equation}
    \theta_i^*(t) = \frac{2\pi q i}{n} + \Omega t
    \label{eq:twisted}
\end{equation}
is an exact solution of Eq.~\eqref{eq:model}. This follows directly from the translational symmetry of the ring: substituting Eq.~\eqref{eq:twisted} into Eq.~\eqref{eq:model}, the phase differences $\theta_j^* - \theta_i^* = 2\pi q(j-i)/n$ depend only on the inter-oscillator distance $j-i$ and not on the absolute index $i$, so the right-hand side reduces to the same constant $\Omega$ for every oscillator, with
\begin{equation}
    \Omega = \bar{\omega}
    + \frac{\sigma}{2r} \sum_{\substack{m=-r \\ m\neq 0}}^{r} \sin\!\left(\frac{2\pi q m}{n}\right)
    + \frac{\sigma_\Delta}{2r(2r-1)} \sum_{\substack{m=-r \\ m\neq 0}}^{r} \sum_{\substack{l=-r \\ l\neq 0,\,l\neq m}}^{r} \sin\!\left(\frac{2\pi q(m+l)}{n}\right).
\end{equation}
Note that the pairwise term vanishes identically since $\sin(2\pi qm/n)$ is odd in $m$, so that $\sum_{m=-r,\,m\neq 0}^{r}\sin(2\pi qm/n) = 0$, and $\Omega$ reduces to
\begin{equation}
    \Omega = \bar{\omega} + \frac{\sigma_\Delta}{2r(2r-1)} \sum_{\substack{m=-r \\ m\neq 0}}^{r} \sum_{\substack{l=-r \\ l\neq 0,\,l\neq m}}^{r} \sin\!\left(\frac{2\pi q(m+l)}{n}\right).
    \label{eq:Omega}
\end{equation}

To assess the linear stability of the twisted state, we introduce a small perturbation $\delta\theta_i$ around the twisted state, writing $\theta_i = \theta_i^* + \delta\theta_i$, and expand Eq.~\eqref{eq:model} to first order in $\delta\theta_i$. The linearized dynamics reads
\begin{equation}
    \delta\dot{\theta}_i = \sum_{j=1}^{n} J_{ij}\,\delta\theta_j,
\end{equation}
where the Jacobian matrix $J$ has off-diagonal elements ($j \in \mathcal{N}_i$)
\begin{equation}
    J_{ij} = \frac{\sigma}{2r}\cos(\theta_j^* - \theta_i^*)
    + \frac{\sigma_\Delta}{2r(2r-1)}\sum_{\substack{k\in\mathcal{N}_i \\ k\neq j}} \cos(\theta_j^* + \theta_k^* - 2\theta_i^*),
    \label{eq:Jij}
\end{equation}
and diagonal elements
\begin{equation}
    J_{ii} = -\sum_{j \in \mathcal{N}_i} J_{ij},
    \label{eq:Jii}
\end{equation}
where Eq.~\eqref{eq:Jii} follows from the rotational symmetry $\theta_i \to \theta_i + C$ of Eq.~\eqref{eq:model}, which guarantees that the uniform perturbation $\delta\theta_i = \text{const}$ is always a zero mode of $J$. Note that the Jacobian depends only on the phase configuration $\theta^*$ and not on the natural frequencies $\omega_i$.

Substituting the twisted state $\theta_j^* - \theta_i^* = 2\pi q(j-i)/n$ into Eqs.~\eqref{eq:Jij}--\eqref{eq:Jii}, the element $J_{ij}$ depends only on the displacement $m = j - i \pmod{n}$ and not on $i$, so $J$ is a circulant matrix whose eigenvalues are given analytically by
\begin{equation}
    \mu_k = \sum_{m=0}^{n-1} J_{0m}\, e^{2\pi\mathrm{i} k m / n}, \quad k = 0, 1, \ldots, n-1.
    \label{eq:eig}
\end{equation}
Excluding the $k=0$ zero mode corresponding to the rotational symmetry, the MTLE is defined as
\begin{equation}
    \lambda_{\max} = \max_{k \neq 0}\, \mathrm{Re}(\mu_k),
    \label{eq:mtle}
\end{equation}
and the twisted state is linearly stable if and only if $\lambda_{\max} < 0$. This analytic structure relies on the exact translational symmetry of the twisted state and does not carry over to the heterogeneous case.

In the presence of frequency heterogeneity ($\omega_{std} > 0$), the exact twisted state Eq.~\eqref{eq:twisted} ceases to be a solution of Eq.~\eqref{eq:model}. Nevertheless, for sufficiently small $\omega_{std}$, the system may still reach a frequency-locked state $\tilde{\theta}_i^*$ in the vicinity of the twisted state, in which all oscillators rotate at a common effective frequency. Such states are identified numerically by the criterion $\mathrm{std}(\omega_i^{\mathrm{eff}}) < 0.02$, where the effective frequency is estimated as
\begin{equation}
    \omega_i^{\mathrm{eff}} = \frac{\tilde{\theta}_i(T_{\mathrm{trans}} + T_{\mathrm{meas}}) - \tilde{\theta}_i(T_{\mathrm{trans}})}{T_{\mathrm{meas}}},
\end{equation}
with $T_{\mathrm{trans}} = 500$ and $T_{\mathrm{meas}} = 100$ time units, consistent with the simulation protocol described in Sec.~\ref{sec:model}. Once a locked state is confirmed, the effect of frequency heterogeneity enters indirectly through the locked phase configuration $\tilde{\theta}_i^*$, which deviates from the exact twisted state and varies across frequency realizations. Substituting $\tilde{\theta}_i^*$ into Eqs.~\eqref{eq:Jij}--\eqref{eq:Jii} yields a Jacobian that is no longer circulant, and the MTLE is computed numerically as the largest real eigenvalue of $J$ after excluding the zero mode.

\begin{figure}[htbp]
\centering
\includegraphics[width=\linewidth]{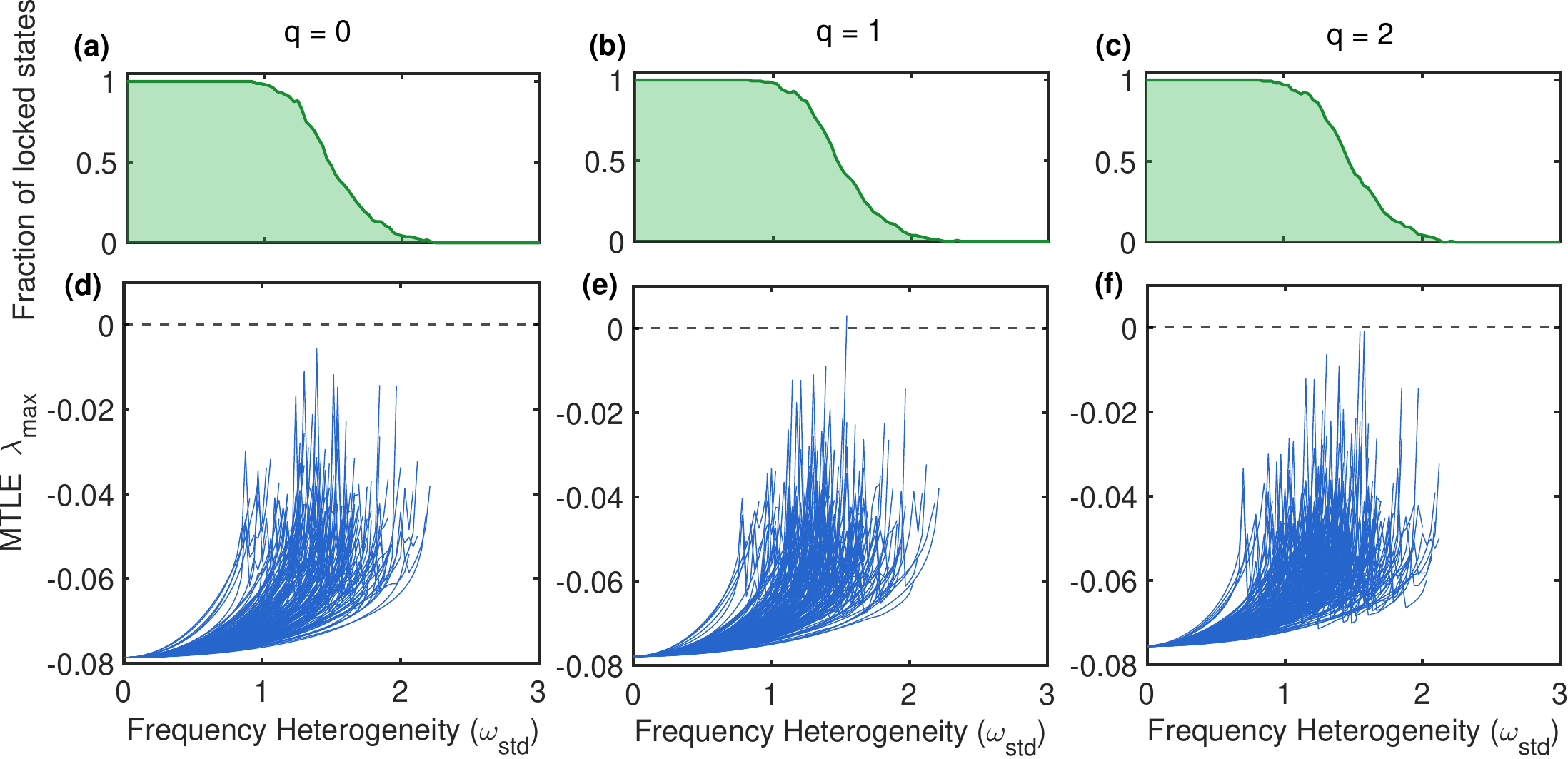}
\caption{Linear stability of frequency-locked twisted states for $q = 0, 1, 2$. Upper panels (a)--(c): fraction of locked states as a function of $\omega_{std}$. Lower panels (d)--(f): MTLE $\lambda_{\max}$ for each of the $160$ frequency realizations; each curve corresponds to one frequency shape tracked across $\omega_{std}$. The dashed line marks $\lambda_{\max} = 0$. Parameters: $n = 83$, $r = 2$, $\sigma = 1.0$, $\sigma_{\Delta} = 5$.}
\label{fig:mtle}
\end{figure}

The results are summarized in Fig.~\ref{fig:mtle}. Two consistent trends emerge across all three winding numbers. First, the fraction of locked states decreases monotonically with $\omega_{std}$: at small $\omega_{std}$, nearly all realizations find a frequency-locked state, whereas as $\omega_{std}$ increases the twisted-state attractor is progressively destroyed, until at large $\omega_{std}$ frequency locking becomes rare. Second, among the realizations that do lock, the MTLE generally increases toward zero as $\omega_{std}$ grows, indicating a progressive weakening of linear stability. All locked states remain linearly stable ($\lambda_{\max} < 0$) throughout, but their stability margin shrinks continuously with increasing heterogeneity.

These two trends together reveal the mechanism by which frequency heterogeneity undermines stability: it simultaneously reduces the probability of frequency locking and erodes the robustness of those locked states that do survive. This stands in direct contrast to the basin stability results of Sec.~\ref{sec:basin}, where moderate frequency heterogeneity was shown to shift the attractor competition 
in favor of more ordered configurations. The two effects thus reflect distinct and competing influences of frequency heterogeneity: a shift in attractor competition favoring more ordered configurations at moderate heterogeneity on one hand, and a weakening of the linear stability of individual locked states on the other. Taken together, they provide a complete mechanistic picture of the dual role of frequency heterogeneity identified in Sec.~\ref{sec:order}.

\section{Discussion and Conclusion}
\label{sec:conclusion}

In this work, we have investigated the effects of frequency heterogeneity on the collective dynamics of a ring network of Kuramoto oscillators with higher-order triadic interactions. Our results reveal a counterintuitive dual role of frequency heterogeneity: it can simultaneously promote global order and undermine the stability of ordered states.

In the strong triadic coupling regime, where chimera and disordered states dominate the homogeneous dynamics, we found that moderate frequency heterogeneity substantially increases the global order parameter $R$, with an optimal heterogeneity strength $\omega_{std}^*$ that grows approximately linearly with $\sigma_{\Delta}$. This order-promoting effect was further elucidated from two complementary perspectives. From the basin stability perspective, moderate frequency heterogeneity enlarges the basin of attraction of the ordered state, redirecting trajectories that would otherwise converge to low-order attractors toward the high-order state. This effect is robust across different initial conditions and frequency realizations, as confirmed by both the ordered fraction analysis and the box plot statistics of $\Delta R$. From the linear stability perspective, however, frequency heterogeneity progressively destroys frequency-locked twisted states and weakens the linear stability of those that survive, as evidenced by the monotonic decrease of the fraction of locked states and the monotonic increase of the MTLE toward zero with growing $\omega_{std}$.

These two competing effects, basin enlargement and linear destabilization, together explain the non-monotonic dependence of $R$ on $\omega_{std}$ observed in Sec.~\ref{sec:order}: moderate heterogeneity shifts the attractor competition toward more ordered 
configurations and thus increases $R$, while excessive heterogeneity destroys the locked states entirely and drives $R$ back toward zero.

Several limitations of the present study point to directions for future work. First, we have focused exclusively on Gaussian frequency distributions; it would be interesting to examine whether the order-promoting effect persists for other distributions, such as Lorentzian or bimodal distributions, which may admit analytical treatment via the Ott--Antonsen ansatz~\cite{OA2008,OA2009}. Second, our linear stability analysis is restricted to twisted states; a systematic study of the stability of higher-$q$ twisted states and their dependence on $\omega_{std}$ remains an open question. Third, the present work considers a simple ring topology with uniform coupling range $r = 2$; extending the analysis to more complex network structures, including random hypergraphs, simplicial complexes, and scale-free networks with higher-order interactions, would shed light on the generality of the observed phenomena. Finally, the interplay between frequency heterogeneity and higher-order interactions in systems with time-varying or adaptive coupling strengths represents another promising direction. We hope that our results will stimulate further investigations into the rich and often counterintuitive dynamics of heterogeneous oscillator networks with nonpairwise interactions.

\begin{acknowledgments}
We thank Wei Wang for helpful discussions. JZ acknowledges the support from the National Natural Science Foundation of China (Grant Nos. 12572037 and 12202195). XL thanks the National Natural Science Foundation of China (Grant No. 12172167) for financial support.
\end{acknowledgments}

\end{document}